\documentclass[useAMS,usenatbib]{mn2e}
\usepackage{graphicx}

\title[V473~Lyrae: a doubly-modulated Cepheid]{V473~Lyrae, a 
unique second-overtone Cepheid with two modulation cycles}
\author[L. Moln\'ar, L. Szabados]{L. Moln\'ar$^{1,2}$\thanks{E-mail:
molnar.laszlo@csfk.mta.hu}, L. Szabados$^{1}$\\
$^{1}$Konkoly Observatory, Research Centre for Astronomy and Earth Sciences\\ 
Konkoly Thege Mikl\'os \'ut 15-17, H-1121 Budapest, Hungary\\
$^{2}$Institute of Mathematics and Physics, Savaria Campus, University of 
West Hungary\\ K\'arolyi G\'asp\'ar t\'er 4, H-9700 Szombathely, Hungary}
\begin{document}

\date{Accepted }

\pagerange{\pageref{firstpage}--\pageref{lastpage}} \pubyear{2014}

\maketitle

\label{firstpage}

\begin{abstract}
V473~Lyrae is the only Galactic Cepheid with confirmed 
periodic amplitude and phase variations similar to the Blazhko effect 
observed in RR~Lyrae stars. We collected all available  
photometric data and some radial velocity measurements to investigate 
the nature of the modulation. The comparison of the photometric and 
radial velocity amplitudes confirmed that the star pulsates in the 
second overtone. The extensive data set, spanning more than 40 years, 
allowed us to detect a secondary modulation cycle with a period of 
approximately 5300 days or 14.5 years. The secondary variations can be 
detected in the period of the primary modulation, as well. 

Phenomenologically, the light variations are analogous to the Blazhko 
effect. To find a physical link, we calculated linear hydrodynamic 
models to search for potential mode resonances that could drive the 
modulation and found two viable half-integer (n:2) and three n:4 
resonances between the second overtone and other modes. If any of 
these resonances will be confirmed by non-linear models, it may 
confirm the mode resonance model, a common mechanism that can 
drive modulations both in RR~Lyrae and Cepheid stars.
\end{abstract}


\section{Introduction}
V473~Lyrae (HR 7308, 19$^{\rm h}$15$^{\rm m}$59\fs49 
+27\degr 55\arcmin 34\farcs 6, J2000.0) is one of the most peculiar 
stars among Cepheids. Its light variations were first noted by 
\citet{breger69}. Independently, \citet{pbt79} identified the star 
as a short-period Cepheid with a tentative period of 3.04 days. 
Shortly after, an extensive radial velocity (RV) measurement series confirmed 
that the true pulsation period is half of the above value, 1.49 days
\citep{bm80a,bm80b}, the shortest value among Galactic Cepheids. 
Burki \& Mayor discovered strong variations in the pulsation amplitude 
over the one and half years of observations. They concluded that the 
star either produced a sporadic or even unique event and it is perhaps 
arriving to the edge the instability strip or, were the variations 
periodic, it shows something similar to the Blazhko effect observed 
in RR~Lyrae stars. 

In the latter case V473~Lyrae would become the first Cepheid to undergo 
periodic amplitude modulations, with a period longer than two years. 
Extensive observations finally revealed that the star indeed shows 
periodic amplitude variations on a time scale of approximately three 
years \citep{pe80}. The modulation period was determined to be 
somewhere around 1210 and 1258 days \citep{breger81,cab91}. 
The origin of the modulation, however, remained a mystery.

Several mechanisms were proposed during the years, but none of them was 
found to be satisfactory. \citet{breger81} proposed the beating of two 
closely spaced pulsation frequencies, the cycle length would, however, 
require a non-radial mode very close to the main mode. \citet{burki84} 
showed that V473~Lyrae has no companions that could cause the amplitude 
variations. Beating and mode interactions were revisited by \citet{vhw95} 
and \citet{breger06}. \citet{auv86} calculated one-zone models with 
variable external flux to explain the modulation but found that the 
modulation period should increase over time. 

\begin{figure*}
\includegraphics[width=175mm]{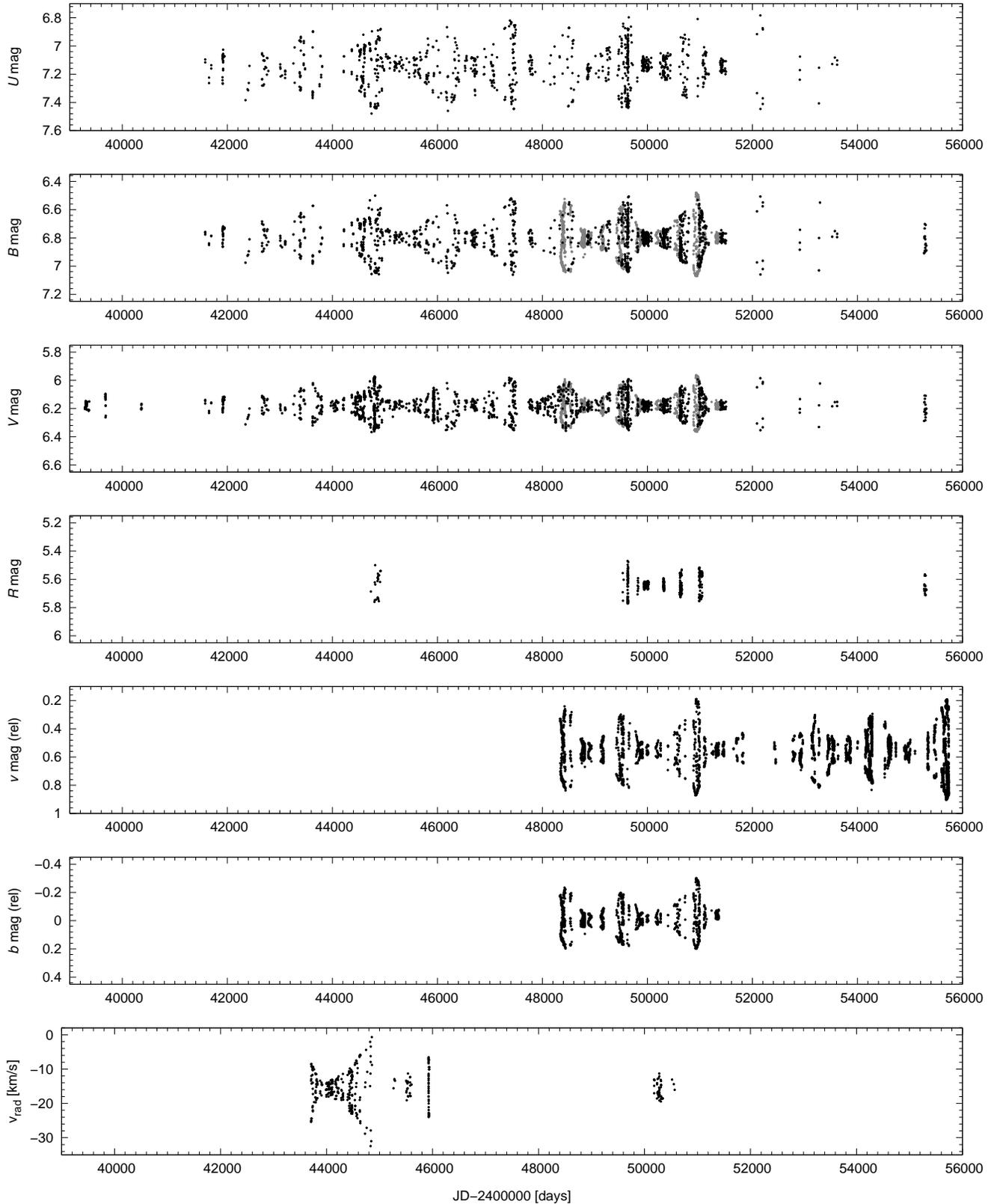}
\caption{Observed variations of V473~Lyrae. First four rows: Johnson 
\textit{UBVR} photometric data. Black: original data; grey: transformed 
from the Str\"omgren colours. The \textit{I} band contains very few 
points hence \textit{I} data have not been plotted. Rows five and six: 
Light curve in Str\"omgren \textit{v} and \textit{b} bands. The 
\textit{u} and \textit{y} bands have the same sampling as the 
\textit{b} band and are omitted too. The last row displays the 
RV variations. The modulation is evident in all cases.}
\label{lc}
\end{figure*}

Finally, based on the apparent similarities to the modulation seen in 
RR~Lyrae stars, the Blazhko effect was considered by several authors. 
Based on the \textit{Hipparcos} light curve, \citet{koen} found that 
the pulsation amplitude is strongly modulated with a period of about 
1200 days but did not detect phase variations. \citet{burki06} did 
find changes in the pulsation phase, but he argued that the star 
exhibits discrete phase jumps instead of the more gradual phase 
modulation. \citet{stothers09} argued that the Blazhko 
effect in RR~Lyrae stars and Cepheids might be driven in a very 
similar manner by turbulent convective cycles. A detailed, direct 
comparison with the RR~Lyrae stars, however, has not been carried out 
so far. Admittedly, since there is no accepted explanation for the 
Blazhko effect yet, the comparison can only be phenomenological. 
But any similarities or differences could provide additional clues 
and constraints for a common mechanism that may operate in both types 
of variables. 

New observational and theoretical developments fundamentally changed 
the way we look at the Blazhko effect in RR~Lyrae stars 
\citep{szabo13rev}. Space-based photometry revealed additional modes 
and resonances in the modulated stars, features that all their 
stable siblings, so far, lack \citep{benko10}. The modulation itself 
was found to be non-repetitive or even extremely variable in several 
stars \citep{gug12}. Moreover, observational findings contradict most 
proposed mechanisms for the Blazhko effect. The most likely explanation 
is currently the mode resonance hypothesis \citep{bk11}. Unlike in the 
non-radial resonant rotator model, the modulation arises from the 
non-linear interaction of the modes instead of the rotation of the star. 

The mode resonance model can be applied to Cepheids too. These stars 
show the same dynamic features that were observed in RR~Lyrae stars: 
multi-mode pulsation, period doubling and strange modes, modes trapped 
close to the surface \citep{mb90,byk97,smolec12}. Therefore we expect 
that mode resonances may occur in these stars as well. However, 
modulated Cepheids are much rarer than modulated RR~Lyrae stars. 
The notable exceptions are the double-mode Cepheids in the Large 
Magellanic Cloud but in those stars both modes show periodic amplitude 
and phase variations \citep{mk09}. Among the Galactic Cepheids, V473~Lyrae 
was the only definitive case, until the recent announcement of a few 
additional stars with indications of Blazhko-like amplitude variability 
by \citet{eg13}.

We started investigating V473~Lyrae to find out if the modulation 
properties are truly similar to the Blazhko effect in RR~Lyrae stars. 
Our first results were promising: we showed that the star exhibits 
simultaneous amplitude and phase modulations \citep{v473first}. In this 
paper we present the results of an in-depth analysis of the 
multicolour photometric data and some limited RV measurements. 
In addition we carried out hydrodynamic simulations to search for mode 
resonances that might be at work in the star.

\section{Data sets}
In order to follow the modulation of V473~Lyrae, we collected all the 
available observations in Johnson colours of the star\footnote{References 
to the observations: \citet{pe80, bm80b, breger81, fernie, bmb82, 
henriksson, af84, eggen, burki86, fabregat, af90, esa, kiss, ignatova, 
breger06, berdnikov08} and references therein; \citet{oja}; Usenko 
(private comm.); Szabados (this paper, Table~\ref{newdata}); AAVSO.}. 
We examined the observations collected by amateur astronomers in the 
AAVSO database: a handful of photometric measurements were included 
but scatter of the visual estimates were too large compared to the 
pulsation amplitude so they were omitted. Most observations were made 
in the \textit{UBV} bands, but we collected a small amount of 
\textit{R-} and \textit{I-}band photometry as well. \textit{UBVR} 
light curves are shown in the first four rows of Fig.~\ref{lc}.

\begin{table}
\caption{$UBV$ photometric data obtained 
at the Konkoly Observatory}
\begin{tabular}{lccc}
\hline
\noalign{\vskip 0.2mm}
JD$_{\odot}$ & $V$ & $B-V$ & $U-B$ \\
2\,400\,000+ & mag & mag & mag \\
\noalign{\vskip 0.2mm}
\hline
\noalign{\vskip 0.2mm}
50252.3594 & 6.175 & 0.609 & 0.263 \\
50281.4811 & 6.193 & 0.614 & 0.329 \\
50609.4014 & 6.178 & 0.621 & 0.343 \\
50633.4003 & 6.220 & 0.658 & 0.316 \\
50634.3635 & 6.073 & 0.566 & 0.295 \\
50956.3977 & 5.994 & 0.530 & 0.280 \\
50957.3853 & 6.328 & 0.661 & 0.362 \\
50960.3980 & 6.314 & 0.661 & 0.303 \\
50961.3615 & 6.249 & 0.675 & 0.300 \\
51051.3351 & 6.203 & 0.612 & 0.303 \\
51052.2981 & 6.232 & 0.649 & 0.334 \\
52086.4270 & 6.031 & 0.558 & 0.208 \\
52087.3843 & 6.288 & 0.663 & 0.264 \\
52150.4565 & 5.967 & 0.517 & 0.180 \\
52151.3126 & 6.336 & 0.701 & 0.290 \\
52195.2715 & 6.005 & 0.548 & 0.202 \\
52196.2477 & 6.315 & 0.683 & 0.294 \\
52197.2143 & 6.253 & 0.686 & 0.313 \\
52198.2225 & 5.994 & 0.532 & 0.238 \\
52902.2449 & 6.211 & 0.650 & 0.259 \\
52904.2392 & 6.185 & 0.626 & 0.241 \\
52906.2333 & 6.115 & 0.605 & 0.236 \\
53266.2687 & 6.313 & 0.695 & 0.280 \\
53267.2642 & 6.159 & 0.619 & 0.257 \\
53286.2548 & 6.004 & 0.523 & 0.208 \\
53518.4131 & 6.166 & 0.603 & 0.240 \\
53569.3638 & 6.136 & 0.593 & 0.234 \\
53612.3055 & 6.164 & 0.608 & 0.242 \\
53614.3085 & 6.135 & 0.613 & 0.235 \\
\noalign{\vskip 0.2mm}
\hline
\end{tabular}
\label{newdata}
\end{table}

\begin{table}
\caption{Available data sets. Columns: data type, photometric colour or 
RV; net number of points; number of points with the 
transformed Str\"omgren data added; time range of the data sets. 
The number of the \textit{uby} data is identical in each band.}
\begin{tabular}{l | c | c | c}
\hline
Type & No.\ of points & With added points & JD range \\
\hline
\textit{U} & 1005 & --   & 2441579-2453614 \\
\textit{B} & 1465 & 2812 & 2441579-2455312 \\
\textit{V} & 2026 & 3369 & 2439289-2455312 \\
\textit{R} &  494 & --   & 2444732-2455312 \\
\textit{I} &  193 & --   & 2444732-2455312 \\
\hline
\textit{v} & 4623 & --   & 2448334-2455741 \\
\textit{uby} & 1347 & -- & 2448334-2451356 \\
\hline
$v_{\rm rad}$            &  376 & -- & 2443706-2450568 \\
\hline
\label{data_points}
\end{tabular}
\end{table}

In addition, we acquired the extensive Str\"omgren-colour observations 
of the Four College APT (Automatic Photoelectric Telescope) run by the 
College of Charleston (SC, USA). The latter data set spans a shorter 
time interval but contains more observations than the Johnson data. 
The first half of the data is covered by \textit{uvby} multicolour 
photometry but the second half was observed in \textit{v}-band only. 
Curiously, the interest in the star declined at the dawn of the 
21st century---at about JD 2451500 or late 1999. Apart from a few 
Johnson measurements, the variations are covered only by the 
Str\"omgren  \textit{v} observations (Fig.~\ref{lc}).

The multicolour part of the Str\"omgren photometry can be converted 
into the Johnson system. According to \citet{williams}, the following 
relation applies to field Cepheids:

\begin{equation}
V = y + 0.028(b-y) - 1.367.
\end{equation}

\citet{cc85} provided additional transformations between the two 
systems. To obtain the \textit{B} colour, we used the following relation:

\begin{equation}
 (B-V) = 0.929 (b-y) + 0.576 (v-b) -0.083.
\end{equation}

We transformed the Str\"omgren photometry into Johnson \textit{B} and 
\textit{V} colours with the help of the above equations. The additional 
points are included in the \textit{B} and \textit{V} light curves in 
Fig.~\ref{lc} with grey colour. The exact number of data points in 
each band is given in Table~\ref{data_points}. The transformed 
points and the observations of \citet{oja} are new additions to the 
\textit{B} and \textit{V} colours compared to \citet{v473first}. 

We also have some RV measurements at our disposal (lowest 
panel in Fig.~\ref{lc}). Unfortunately, we could not access the 
extensive measurements shown in \citet{burki06}. Nevertheless, 
the limited data set covers most modulation phases.

\section{Identification of the pulsation mode}
\label{id}
Being a unique Cepheid in our Galaxy, the pulsation mode of V473~Lyr 
is a key issue for explaining its peculiar behaviour. Presence of the 
Blazhko modulation was reported for a number of Cepheid variables in 
the Large Magellanic Cloud. Such modulation phenomenon only occurs 
among double-mode Cepheids pulsating simultaneously in the first and 
second overtone modes \citep{mk09}.

The earliest discussions on the pulsation mode of V473~Lyr are found 
in the papers by \citet{bmb82} and \citet{burki86}. In the first paper, 
they determined the pulsation constant, $Q$, from the pulsation period, 
mass, and radius of the star, and compared with the theoretical $Q$ 
value obtained for various modes of oscillation. Their conclusion is 
that pulsation in the second radial overtone is most probable in the 
case of V473~Lyr but the first overtone cannot be excluded, either. 
In their subsequent paper, \citet{burki86} derived physical properties 
of V473~Lyr using three versions of the Baade-Wesselink method, and 
based on the stellar radius obtained they again concluded that this 
Cepheid most probably pulsates in the second or higher overtone. 
The position of V473~Lyr on the period-luminosity plot is also 
compatible with the overtone pulsation \citep*{fabregat}.

There are, however, other methods for mode identification of Cepheids 
not applied in the case of V473~Lyr so far. These methods are based 
on various phenomenological properties (amplitude, shape, phase relations) 
of the phase curves of the brightness and RV variations.

The most commonly used method of determining the mode of pulsation 
for a Cepheid is based on the Fourier decomposition of the photometric 
phase curve. The Fourier coefficients defined by Simon \& Lee (1981) 
show a characteristic progression as a function of the pulsation period.
Based on the $R_{21}$ and $R_{31}$ amplitude parameters and $\phi_{21}$ 
and $\phi_{31}$ phase parameters the overtone pulsators are clearly 
separated from Cepheids pulsating in the fundamental mode (for Cepheids 
in the Small Magellanic Cloud see Figures 2-3 in \citealt{sosz10}). 
However, Cepheids pulsating in the first two overtone modes are partly 
overlap in the Fourier parameter vs.\ pulsation period diagrams. 
In the case of Magellanic Cepheids, the distinction between first- and 
second-overtone pulsators can be made by their position in the period 
vs. apparent brightness (or Wesenheit magnitude) graphs, because all 
Cepheids situated in either Magellanic Cloud are practically at the 
same distance from us. This is not a viable alternative for Galactic 
Cepheids because they are at various distances.

\begin{figure}
\includegraphics[width=84mm]{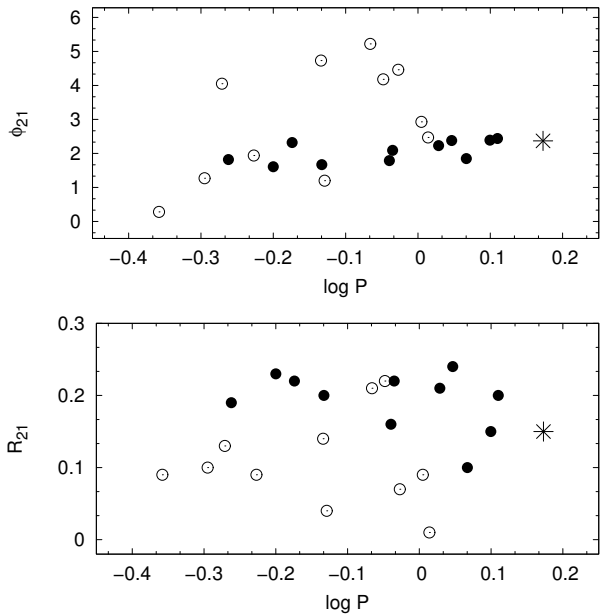}
\caption{Fourier parameter relations for Galactic double-mode Cepheids 
versus the logarithm of the pulsation mode period in days. Top panel: 
the $\phi_{21}=\phi_2-2\phi_1$ phase relation; bottom panel the 
$R_{21}=A_2/A_1$ amplitude relation. Filled points: first overtone; 
circles: second overtone. The star marks the position of V473~Lyrae. 
Because of the overlap and the large scatter of the second-overtone 
points, Fourier parameters alone cannot identify the pulsation mode 
of V473~Lyrae. }
\label{fig-r21-ph21}
\end{figure}

The available Fourier parameter vs.\ period diagrams for Galactic 
Cepheids only involve fundamental and first-overtone pulsators and 
no single-period, second-overtone Galactic Cepheid is known with 
certainty. There are, however, 11 known double-mode Cepheids in our 
Galaxy which pulsate simultaneously in first and second overtones 
(V363~Cas, V1048~Cen, V767~Sgr, ASAS 062735+0949.8, ASAS 064135+0756.6, 
ASAS 074343-2050.3, ASAS 083434-4134.6, ASAS 131714-6605.0, 
ASAS 142346-5829.4, ASAS 191351+0251.3, and NSVS 2030690). To perform 
a Fourier decomposition of their light curves, the photometric data 
available in the online data bases ASAS \citep{pojmanski} and NSVS 
\citep{wozniak} have been analyzed. The Fourier decomposition was 
performed using the program package MUFRAN (Csubry \& Koll\'ath 2004). The 
frequency values of the first and second overtones, $f_1$ and $f_2$,
were taken from the literature, and the following harmonics
and linear combinations of these two basic frequencies
were taken into account in the fitting procedure:
$f_1$, $2f_1$, $3f_1$, $f_2$, $2f_2$, $f_1+f_2$,
$2(f_1+f_2)$, $f_2-f_1$, $2f_1+f_2$. Inclusion of
further frequency components would have been meaningless
in view of their negligible amplitudes and the quality of
the photometric data sets. The frequencies listed above were
fitted simultaneously, and than the $R_{21}$ and $\phi_{21}$
parameters were derived. The resulting $R_{21}$ vs. $\log P$ 
and $\phi_{21}$ vs. $\log P$ diagrams are shown in Fig.~\ref{fig-r21-ph21}. 
In these figures filled circles denote the respective Fourier parameter 
for the 1st overtone of the Galactic beat Cepheids, while open circles 
represent the corresponding value for the 2nd overtone oscillations. 
A star symbol shows the position of V473~Lyrae in both diagrams. 
Unfortunately, the quality of the available photometric data does not 
permit a reliable determination of the $R_{31}$ and $\phi_{31}$ 
parameters for the Galactic double-mode Cepheids.

\begin{figure}
\includegraphics[width=84mm]{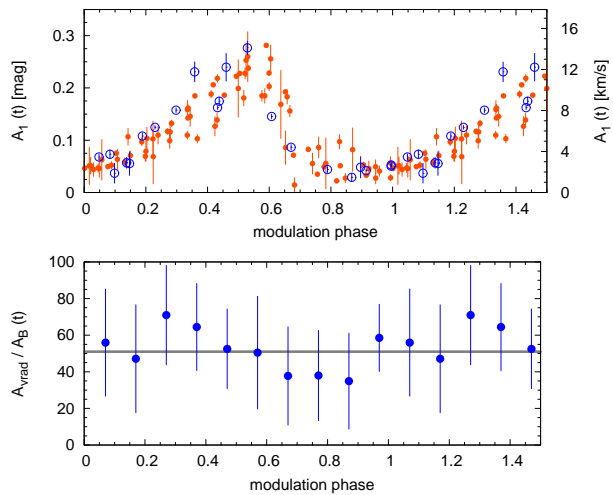}
\caption{Upper panel: the amplitude variations of the \textit{B} 
passband (orange dots) and RV data (blue circles), 
scaled together and folded with the primary modulation period. 
Lower panel: the $q=A_{V_{\rm rad}} / A_B$ scaling factor, calculated 
for 10 phase bins. The average (grey line) is 51, indicating that the 
star pulsates in the second overtone. }
\label{q_parameter}
\end{figure}

Figure~\ref{fig-r21-ph21} indicates partial overlapping of the Fourier 
parameters for the first and second overtone pulsations in Galactic 
Cepheids, similarly to the case of Magellanic Cepheids. The position 
of V473~Lyrae in these diagrams does not prefer either overtone mode, 
though excludes fundamental-mode pulsation.

There are two methods of mode determination utilizing both photometric 
and RV measurements. On the one hand, the first-order 
phase lag between the decomposed RV and photometric curves 
($\phi_{21}^{V_{\rm r}} - \phi_{21}^{\rm mag}$) is a good indicator of 
the pulsation mode, at least for differentiating between fundamental-mode 
and 1st overtone Cepheids \citep{ogloza}. This method is only applicable 
for pulsation periods longer than 3.5~days, so useless in the case 
of V473~Lyrae. On the other hand, the ratio of the RV 
and photometric amplitudes also hints at the mode of pulsation 
\citep{balstob79}. The ratio of these amplitudes, 
$q = A_{V_{\rm rad}}/A_B$, is about $30-33$ for fundamental-mode 
Cepheids and $q \approx 35-40$ for Cepheids oscillating in the first 
overtone \citep{klagy09}. For higher overtones $q$ is even larger. 
The $q$ parameter is about $51\pm10$ for V473~Lyrae, (the largest 
value among Galactic Cepheids), although the modulation and the 
sparse sampling makes the determination somewhat uncertain 
(Fig.~\ref{q_parameter} and Sect.~\ref{vradcomp}). The ratio is the 
same for the high and the low-amplitude phases of the long cycle. The 
unusually high value is a clear evidence that the star is indeed 
pulsating in the second overtone. We note that \citet{klagy09} warned 
that binarity could increase the value of the $q$ ratio but the constant 
time-averaged RV value \citep{burki06} indicates no 
close-by, high-mass companions.

\section{Fourier analysis}
\begin{figure}
\includegraphics[width=84mm]{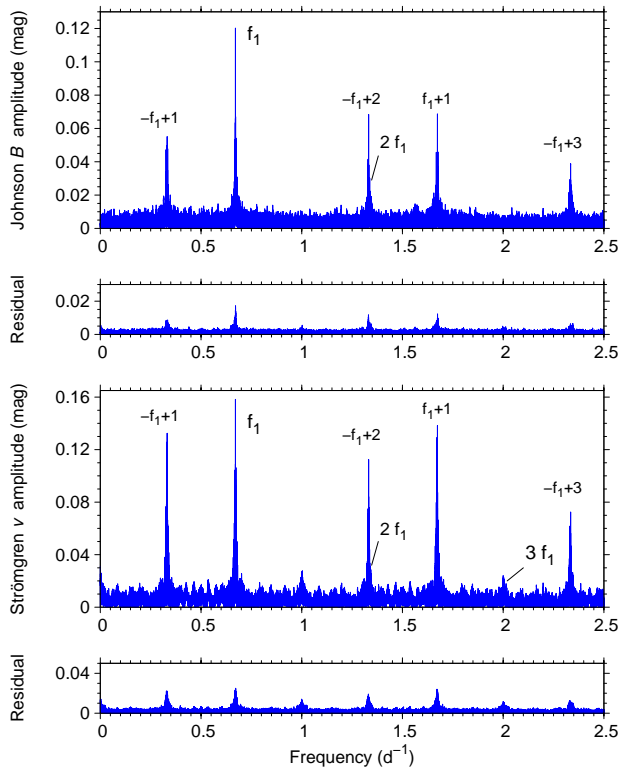}
\caption{Fourier spectra of the Johnson \textit{B} and Str\"omgren 
\textit{v} data sets of V473~Lyrae. Real and alias peaks are indicated. 
The bottom panels show the residual spectra after prewhitening with 
the frequencies indicated in Fig.~\ref{echelle}. }
\label{fourier}
\end{figure}

We investigated the Fourier spectra of all data sets. Spectra from the 
two best colours, the Johnson \textit{B} and Str\"omgren \textit{v} 
are plotted in Fig.~\ref{fourier}. Since the pulsation frequency is 
almost $2/3$ d$^{-1}$, ground-based observations of this star suffer 
from significant aliasing. Daily aliases of the $-f_1$ value from the 
negative side of the spectrum leak into the positive side at 
$-f_1+1$ d$^{-1}$, $-f_1+2$ d$^{-1}$, etc.\ values, among which the 
$-f_1+2$ d$^{-1}$ value coincides with the $2f_1$ value, as indicated 
in Fig.~\ref{fourier}. Similarly, the $3f_1$ peak coincides with the 
3 d$^{-1}$ alias peak. The Str\"omgren data collected from a 
single site, suffers more from these problems.

Cepheid light curves are much more sinusoidal than the variations of 
RR~Lyrae stars so the amplitudes of the successive harmonics (integer 
multiplets of the main frequency) decrease very fast. The first harmonic 
($2f_1$) was identified in all but the very sparse \textit{I} data sets. 
However, the $3f_1$ peak (or a corresponding side-peak) was only
detectable in the \textit{B, V} and \textit{v} bands that provide the most 
extensive temporal coverage. The low-frequency modulation peak ($f_m$) 
was identified in most bands, although the exact frequency values were 
uncertain for the shorter \textit{uby} data sets. The Fourier parameters 
of the \textit{B} band data are summarized in Table \ref{b_freq_table}.

\subsection{Modulation side-peaks}
Amplitude and/or phase variations introduce modulation side-peaks around 
the main peaks in the frequency spectrum. The exact number and amplitudes 
of these peaks depend not only on the detection limit but on the details 
of the modulation process as well. On the one hand, \citet{szj09} showed 
that a sinusoidal signal modulated both in amplitude and phase with a 
periodic function (e.g.\ another sine) produces an infinite series of 
side-peaks with decreasing amplitudes. On the other hand, non-sinusoidal 
modulation, especially in the pulsation phase, may create more complicated 
patterns around the main peaks \citep*{benko11}. Hence the number of 
side-peaks and the amplitude ratios provide valuable information about 
the nature of the modulation.

\begin{figure}
\includegraphics[width=84mm]{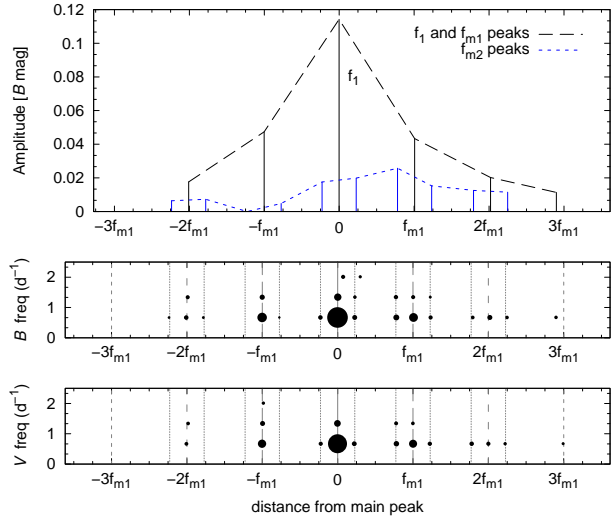}
\caption{Summary of the detected main frequency peaks and side-peaks, 
except for the modulation frequencies themselves. Top panel: identified 
peaks around the pulsation frequency. Black: $f_1$ and the primary 
modulation peaks. Blue: secondary modulation peaks and combination 
peaks. Numerical values are summarized in Table~\ref{b_freq_table}. 
Lower panels: the peaks are plotted similar to an echelle diagram: 
successive orders of harmonics with the corresponding side-peaks were 
placed above each other. Point sizes are derived from the amplitudes 
but the second and third rows of points in both panels are enlarged for 
better visibility.}
\label{echelle}
\end{figure}

\begin{figure}
\includegraphics[width=84mm]{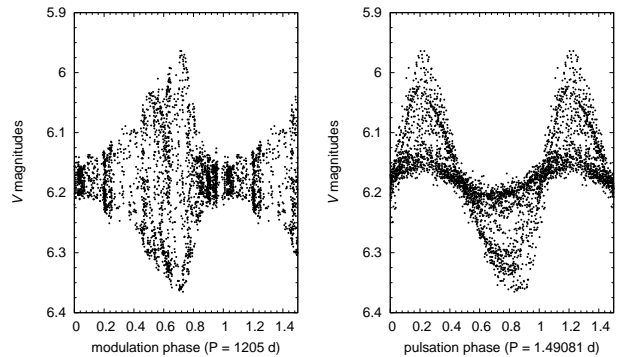}
\caption{Johnson \textit{V} band light curves, folded with the 
modulation and pulsation periods. The modulation is not symmetric: 
the pulsation amplitude decreases much faster than it increases. 
Because of the additional variations (the secondary modulation, see 
Sect.~\ref{sec_longmod}), we only used data between JD~2447700 and 
2451500 in this plot.}
\label{folded}
\end{figure}

\begin{table}
\caption{Fourier solution of the Johnson \textit{B} data set. We used 
the Period04 and MuFrAn tools throughout our analysis \citep{period,mufran}. 
Uncertainties were calculated with the Monte--Carlo simulation of Period04. 
The original \textit{B}-band spectrum and the residual after prewhitening 
with the main peaks and primary side-peaks can be seen in Fig.~\ref{fourier}.}
\begin{tabular}{l@{\hskip2mm}c@{\hskip2mm}c@{\hskip2mm}c@{\hskip2mm}c@{\hskip2mm}c@{\hskip2mm}c}
\hline
Peak & $f$ & $A$ & $\phi$ & $ \sigma(f)$ & $\sigma(A)$ & $\sigma(\phi)$ \\ 
     & d$^{-1}$ & mag & $\frac{{\rm rad}}{2\pi}$ & d$^{-1}$ & mag & $\frac{{\rm rad}}{2\pi}$\\ 
\hline
\multicolumn{7}{l}{~}	 \\
\multicolumn{7}{c}{\textit{Main peaks}}	 \\[2ex]	
$f_{m1}$ & 0.000856 & 0.009 & 0.779 & 0.000018 & 0.010 & 0.029 \\
$f_1$	 & 0.670784 & 0.114 & 0.060 & 0.000003 & 0.005 & 0.007 \\
$2f_1$	 & 1.341571 & 0.017 & 0.333 & 0.000029 & 0.004 & 0.040 \\
$3f_1$	 & 2.01242  & 0.005 & 0.75  & 0.00098  & 0.003 & 0.15 \\
\multicolumn{7}{l}{~}	 \\
\multicolumn{7}{c}{\textit{Side-peaks of the primary modulation}} \\[2ex]
$f_1$$-$$2f_{m1}$  & 0.66912  & 0.018 & 0.83  & 0.00060  & 0.011 & 0.23 \\
$f_1$$-$$f_{m1}$   & 0.669954 & 0.047 & 0.235 & 0.000005 & 0.004 & 0.015 \\
$f_1$+$f_{m1}$   & 0.671620 & 0.043 & 0.532 & 0.000063 & 0.012 & 0.043 \\
$f_1$+$2f_{m1}$  & 0.672462 & 0.020 & 0.633 & 0.000012 & 0.006 & 0.068 \\
$f_1$+$3f_{m1}$  & 0.67319  & 0.011 & 0.52  & 0.00173  & 0.008 & 0.36 \\
$2f_1$$-$$2f_{m1}$ & 1.339919 & 0.007 & 0.404 & 0.000019 & 0.004 & 0.052 \\
$2f_1$$-$$f_{m1}$  & 1.34074  & 0.011 & 0.60  & 0.00424  & 0.013 & 0.28 \\
$2f_1$+$f_{m1}$  & 1.34240  & 0.007 & 0.01  & 0.00049  & 0.173 & 0.17 \\
$3f_1$+$f_{m1}$  & 2.01260  & 0.003 & 0.80  & 0.02004  & 0.003 & 0.34 \\
\multicolumn{7}{l}{~}	 \\
\multicolumn{7}{c}{\textit{Side-peaks of the secondary modulation}} \\[2ex]
$f_1$$-$$f_{m2}$ &	0.670595 & 0.018 & 0.566 & 0.000019 & 0.004 & 0.071 \\
$f_1$+$f_{m2}$ &	0.670973 & 0.020 & 0.062 & 0.000014 & 0.004 & 0.039 \\
$2f_1$+$f_{m2}$ &	1.34176  & 0.006 & 0.34  & 0.00026  & 0.003 & 0.17 \\
\multicolumn{7}{l}{~}	 \\
\multicolumn{7}{c}{\textit{Combination peaks}}	\\[2ex]
$f_1$$-$$2f_{m1}$$-$$f_{m2}$ & 0.66893  & 0.006 & 0.86  & 0.00242  & 0.011 & 0.18 \\
$f_1$$-$$2f_{m1}$+$f_{m2}$ & 0.66931  & 0.007 & 0.136 & 0.00073  & 0.007 & 0.083 \\
$f_1$$-$$f_{m1}$+$f_{m2}$  & 0.67014  & 0.005 & 0.456 & 0.00007  & 0.004 & 0.092 \\
$f_1$+$f_{m1}$-$f_{m2}$  & 0.67143  & 0.026 & 0.15  & 0.00181  & 0.025 & 0.24 \\
$f_1$+$f_{m1}$+$f_{m2}$  & 0.671809 & 0.015 & 0.486 & 0.000031 & 0.021 & 0.071 \\
$f_1$+$2f_{m1}$$-$$f_{m2}$ & 0.672273 & 0.013 & 0.319 & 0.000020 & 0.011 & 0.046 \\
$f_1$+$2f_{m1}$+$f_{m2}$ & 0.672650 & 0.012 & 0.565 & 0.000046 & 0.006 & 0.059 \\
$2f_1$+$f_{m1}$$-$$f_{m2}$ & 1.34221  & 0.008 & 0.67  & 0.00218  & 0.005 & 0.33 \\
$2f_1$+$f_{m1}$+$f_{m2}$ & 1.34259  & 0.004 & 0.97  & 0.01554  & 0.171 & 0.39 \\
\label{b_freq_table}
\end{tabular}
\end{table}

Previous studies only identified triplet components, 
\textit{e.g.}\ $f_1 \pm f_m$ side-peaks \citep{koen, breger06}. 
We detected a single quintuplet peak during the preliminary analysis 
\citep{v473first}. We repeated the analysis with the extended data sets. 
After a thorough search we identified side-peaks up to the third order 
($f_1\pm 3f_m$) in different bands. We determined the modulation period 
to be $P_m = 1205 \pm 3$ days, based on the individual modulation 
frequencies. 

In Fig.~\ref{folded} we folded the light curves with the modulation 
and pulsation periods, respectively. The strong amplitude variations 
can be clearly seen. Interestingly, the modulation cycle is very 
asymmetric (left panel of Fig.~\ref{folded}): the increase of the 
amplitude lasts much longer than the fast decrease which is then followed 
by an extended phase in amplitude minimum. In contrast, the Blazhko 
effect in RR~Lyrae stars is relatively symmetric, although some stars 
do show some asymmetry \citep{benko10, skarka14}. 

\begin{figure}
\includegraphics[width=84mm]{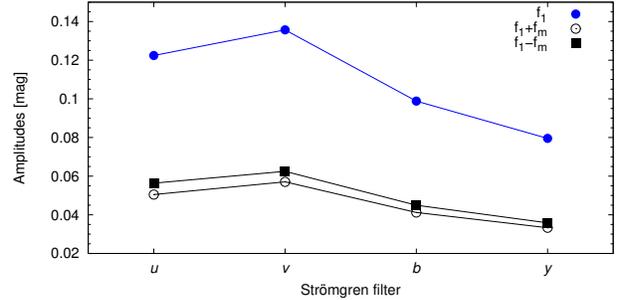}
\caption{Amplitudes of the $f_1$ main pulsation peak and the 
$f_1\pm f_m$ triplet modulation side-peaks in each Str\"omgren band. 
The ratio of the pulsation and modulation peaks is the same in each 
bandpass. }
\label{peaks}
\end{figure}

We also checked the colour dependence of the ratio of the pulsation and 
modulation peaks. We used the multicolour part of the Str\"omgren data 
set. The ratio turned out to be constant in all four passbands so the 
modulation does not exhibit any colour dependence (Fig.~\ref{peaks}).

\subsection{Residual power}
\label{comb_freq}
Interestingly, significant power remained in the spectra even after 
removing all side-peaks (lower panels of Fig.~\ref{fourier}). One can 
of course continue to prewhiten those structures: overlapping multiplet 
structures were already observed in RR~Lyrae stars with multiple or 
changing modulation cycles \citep{czlac, gug12}. \citet{cab91} already 
reported that the modulation seems not to be perfectly periodic. 
Further inspection revealed a regular structure of additional side-lobes 
around some of the main peaks and primary modulation side-peaks. 
Such pattern can be interpreted as a secondary modulation superimposed 
on the first one. Most additional peaks are only slightly farther from the 
primary modulation peaks than the first side-peak of the window function, 
\textit{viz.}\ $f_{m2} \sim 0.00018-0.00019\,{\rm d^{-1}}$ compared to 
$f_{\rm wpeak}=0.00014-0.00017\, {\rm d^{-1}}$. The $f_2$ frequency we 
identified in the preliminary analysis is also part of this structure 
\citep{v473first}. We concluded that the remaining power comes from 
additional variations with time scales comparable to the length of 
the data set. The average distance of the secondary side-peaks is 
$f_{m2}=0.0001890\pm 0.0000035 \,{\rm d^{-1}}$, corresponding to a 
period of $5290\pm 96$ days. The detected peaks in the Johnson \textit{B} 
and \textit{V} colours are summarized in Fig.~\ref{echelle}. 
Note that although the primary modulation peaks form a relatively 
symmetric structure, the amplitudes of the $f_{m2}$ do not. The highest 
peak appears next to the $f_1+f_{m1}$ frequency while we found the 
lowest peaks (with one non-detection) around the $f_1-f_{m1}$ frequency. 
The frequencies derived from the Johnson \textit{B} data set that 
provided the highest signal-to-noise ratios, are summarized in 
Table~\ref{b_freq_table}.

\begin{figure*}
\includegraphics[width=175mm]{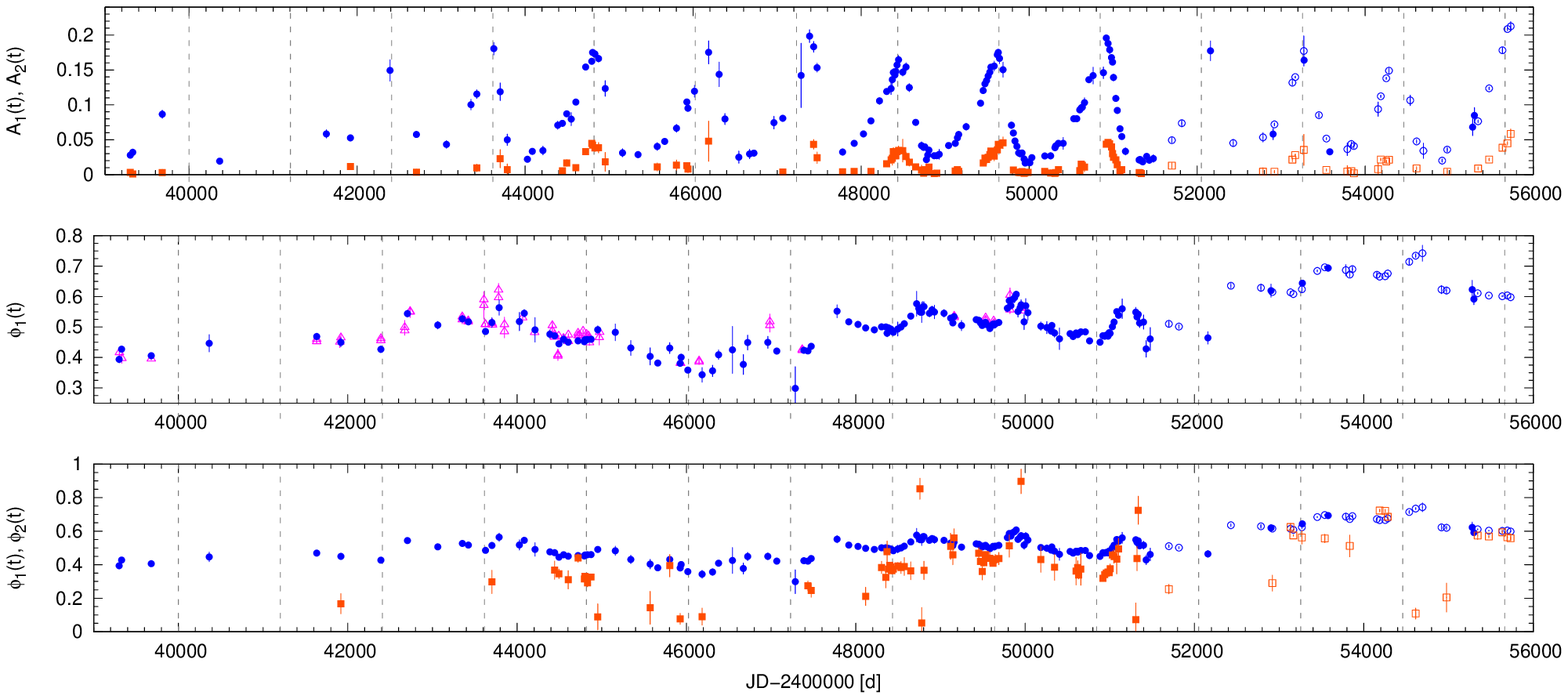}
\caption{Amplitude and phase variations of V473~Lyrae. Top panel: 
amplitudes of the $f_1=0.670775\, {\rm d^{-1}}$ (blue dots) and 
$2f_1$ (orange squares) peaks. Middle panel: phases of the $f_1$ 
peaks (blue), purple triangles are from \citet{berdpast94} and 
\citet{berdnikov97}. Units are rad/2$\pi$. Lower panel: phases of both frequency peaks. 
In the case of $\phi_1$, points with $\Delta \phi_1 < 0.2$ 
uncertainty are only shown. Filled and empty points correspond to 
Johnson \textit{V} and Str\"omgren \textit{v} colours, respectively, 
\textit{v} amplitudes were scaled down to match the \textit{V} 
amplitudes. Grey vertical lines mark the modulation cycles with a 
spacing of 1205 days. }
\label{amp_phase} 
\end{figure*}

\section{Amplitude and phase variations} 
The large residuals in the Fourier spectra indicate that the 
modulation of the star may be variable. Preliminary analysis 
indicated that the star indeed shows changes in the modulation 
\citep{v473first}. We divided the data sets into bins with lengths 
between 10 and 100 days, depending on the distribution of the points. 
Although the pulsation amplitude is higher in the \textit{B} band, 
we used the Johnson \textit{V} data for this analysis because of its 
better temporal coverage. The pulsation frequency was set to 
$f_1=0.670775\, {\rm d^{-1}}$, adopted from the ephemeris used by 
\citet{berdpast94}. If the temporal coverage allowed, we fitted the 
amplitude and phase ($A_2$, $\phi_2$) of the $2f_1$ frequency as well. 
However, the $\phi_2$ value was very uncertain in all but the 
highest-amplitude phases, therefore we could not calculate 
the epoch-independent relative Fourier-phases, $\phi_{21}$ and $\phi_{31}$ 
\citep{phi21} for any other parts of the light curve. The $\phi_{3}$ 
parameter was barely detectable for ther entire data set. Because of 
this, we resorted to use $\phi_1$ itself in the analysis. The 
results are shown in Fig.~\ref{amp_phase}. We derived the 
\textit{B}-band and RV amplitudes in the same way to identify 
the mode of the pulsation in Sect.~\ref{id}.

It is clear that both the amplitude and the phase are modulated 
with the $P_m = 1205 \pm 3$ day period. The variations of the 
amplitude repeat fairly regularly with slight changes in the 
maximum modulation amplitude. On the other hand, the phase shows 
much more complex changes. Significant additional variations occur 
on about $2-4$ modulation-cycle-long time scales.

\subsection{Evidence of a secondary modulation cycle}
\label{sec_longmod}
Although the Blazhko-like modulation in a Cepheid is in itself a 
unique feature, the light variations of the star turned out to be 
even more complex. The residual power in the Fourier spectra 
(Fig.~\ref{fourier}) and the long-term changes in the phase 
variation curve (Fig.~\ref{amp_phase}) already indicated that 
additional variation may be present in the star. The times of maximum 
pulsational amplitude also seem to deviate from the predicted times 
in the top panel of Fig.~\ref{amp_phase}. 

\begin{figure}
\includegraphics[width=84mm]{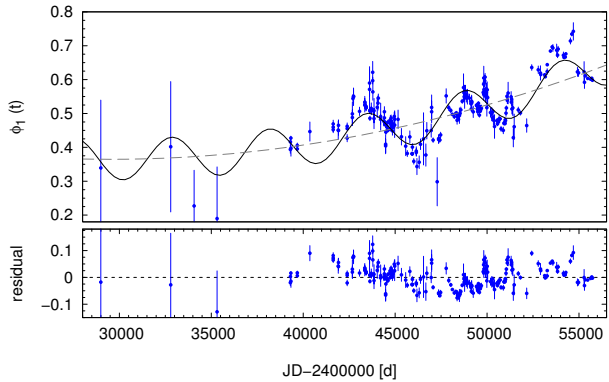}
\caption{The phase variations fitted with a quadratic term, 
indicating a shortening pulsation period (the phase variation is
the negative of the $O-C$ values), and a sinusoidal with a period 
of 5290 days. The longer modulation cycle is definitely present 
in the phase variations. A few early, uncertain points by 
\citet{berdnikov97} are also included. }
\label{phase_longmod}
\end{figure}

Further investigation of the phase variation revealed two 
additional variations (Fig.~\ref{phase_longmod}). A secular 
period change of 
$\dot{P}=-0.01673\pm0.00058$ s/yr $= (-5.30\pm 0.18) \cdot10^{-10}$~s/s 
is present, indicating evolutionary effects. Note that the phase 
variations have opposite sign compared to the $O-C$ values hence 
the quadratic term in Fig.~\ref{phase_longmod} corresponds to a 
decrease in the average pulsation period.

The second term in Fig.~\ref{phase_longmod} is a sinusoidal 
function with a period of $P_{m2}=5290$ d, the value we identified
in the Fourier spectra. The amplitude of the variation is 
$A_\phi(t)=0.061(\pm0.004)\, {\rm rad}/2\pi$ or $0.091(\pm0.006)$ days. 
This long-term modulation is clearly different from the faster primary 
modulation cycle that can be clearly seen in the lower panel of 
Fig.~\ref{phase_longmod} after removing the two longer variations. 

\section{Discussion}

\subsection{Origin of the secondary modulation}
A periodic phase modulation could arise from the light-time effect
if the star is a member of a binary system. 
We calculated the approximate mass limit of a companion from the 
well-known binary mass function:
\begin{equation}
f(m)=4 \pi^2 \, \frac{(A\,c\, \mathrm{sin}\,i)^3}{G\,P^2} = \frac{(M_2\,\mathrm{sin\,}i)^3}{(M_1+M_2)^2}
\end{equation}
where $A$ is the semi-amplitude of the $O-C$ or phase variations, 
$c$ is the speed of light, $M_1$ and $M_2$ are the masses of the 
stars, $P$ is the orbital period, $G$ is the gravitational constant 
and $i$ is the inclination of the orbit. By setting $P=5290$~d, 
$A=0.091$ d and $M_1 = 4\, M_\odot$ \citep{fabregat}, the calculation 
yields a mass limit of $M_2\,\mathrm{sin}i = 24\, M_\odot$ for a 
hypothetical companion. However, the delay and the modulation 
period would mean a RV amplitude of $v_{\rm rad}=32\, 
{\rm km/s}$. Neither the last panel in Fig.~\ref{lc} nor the more 
extended observations of \citet{burki06} show such large variations 
in the RV measurements. Hence, binarity can be ruled out.

\begin{figure}
\includegraphics[width=84mm]{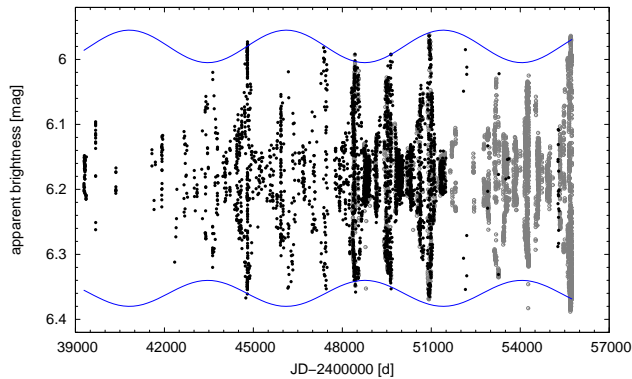}
\caption{Approximate envelope of the secondary modulation. 
Black points are the Johnson \textit{V}, grey circles are the 
scaled Str\"omgren \textit{v} data points. The modulation period 
is 5290 days (14.5 years). }
\label{longmod}
\end{figure}

As the secondary phase modulation appeared to be an intrinsic 
variation of the star, we revisited the light curves to look for 
corresponding amplitude changes. In Fig.~\ref{longmod} we applied 
an envelope to the Johnson \textit{V} data with the same period 
and phase as that of the secondary phase modulation. The 
Str\"omgren \textit{v} data were scaled to the \textit{V} data 
to better visualize the effect. Although the highest-amplitude 
phases of the primary modulation are missing in several cycles, 
the envelope curves clearly fit the observations, indicating the 
presence of the secondary modulation. The peak-to-peak amplitude 
of the variation is approximately $A_{m2} = 0.09 \pm 0.04$ magnitudes 
in \textit{V} colour.  

Multiple modulation cycles have been already known to exist in 
some Blazhko RR~Lyrae stars (see Sect.~\ref{multi}). 
The primary and secondary cycles can be close to each other, 
as in the case of CZ Lac where the two cycles were observed close 
to the 5:4, and in the next season to the 4:3 period ratios 
\citep{czlac}. The period ratio of the two variations is 
$P_{m2}/P_{m1} = 4.4 \pm 0.15$. The uncertainties can accommodate 
an exact $P_{m2}/P_{m1} =$ 9:2 ratio but it will take further, 
long-term monitoring to determine the modulation periods more 
accurately. This aspect of the modulation of V473~Lyr agrees with 
the Blazhko effect.

\subsection{Relation of the two modulation cycles}
Another interesting question is whether the secondary modulation 
is independent of the primary and affects only the pulsation or 
the two modulations interact with each other. To answer this, 
we derived the $O-C$ curves of the modulation. 
For the phase variation we calculated the times of the phase 
maxima which are relatively sharp features. In the case of the 
amplitude variations, we calculated both the amplitude maxima and 
the times when the $A(t)$ curves crossed the 0.1-magnitude level 
(the approximate mean light level) upwards. The latter, arbitrary 
point on the rising branch can be measured more accurately than the maxima. 

\begin{figure}
\includegraphics[width=84mm]{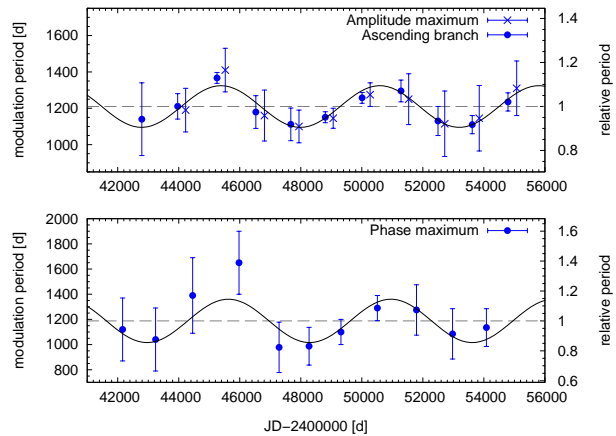}
\caption{Variations in the cycle length of the primary modulation. 
Top: based on the $O-C$ variations of the amplitude modulation data 
(maxima and mean lights of the ascending branch). Bottom: based on 
the phase modulation data. }
\label{modper}
\end{figure}

All three measures of the cycle length of the primary modulation 
in Fig.~\ref{modper} indicate that the modulation period is 
indeed changing. Although the $O-C$ variation of the phase and 
amplitude maxima is marginal due to the large uncertainties, the 
phase and period agrees with the secondary modulation we detected 
in the phase variations of the pulsation (Fig.~\ref{phase_longmod}). 
The mean brightness levels on the rising branch on the other hand 
can be determined more accurately and fit to the variations as well. 
The amplitudes are somewhat different: the times of the amplitude 
maxima vary by $115 \pm 12$ days (about 9.5\%) whereas the times 
of the phase maxima vary by $ 172 \pm 51$ days or about 14\%. 
The average period of the primary modulation was found to be 
$P_{m1}(A(t))= 1209.7 \pm 8.2$~d and $P_{m1}(\phi(t)) = 1188 \pm 35$~d, 
in agreement with the results of the Fourier analysis. If we allow 
the secondary modulation period to be fitted, the amplitude data 
results in a period of $P_{m2}=5330 \pm 137$ days. The phase data 
provide only a very uncertain period.

The variations of the primary modulation period indicate that the 
modulations may possibly interact with each other. \citet{benko11} 
argue that two parallel, independent modulating waves are unlikely 
to occur in real stars without interaction. According to their 
mathematical formulation, in the case of an amplitude and/or 
frequency modulation cascade, linear combinations of the two 
modulation frequencies should be present at the low-frequency end 
of the Fourier spectrum. However, the data is insufficient to 
detect $f_{m2}$ itself or its combinations with $f_{m1}$. 
Unfortunately, the combination peaks between the side-peaks can 
arise simply from the presence of the two frequency modulations 
and their presence cannot distinguish between independent and interacting modulations. 
The current time span of the data covers only two and a half 
secondary modulation cycles, therefore further long-term monitoring 
will be necessary to detect the signs of any interaction.

The periodic variations in the length of the primary modulation 
cycle may explain the different values found in the literature: 
e.g., \citet{cab91} determined $P_{m1}$ to be 1258 days but his 
data covered mostly the time interval of significantly longer 
modulation period around JD 2445000. 

\begin{figure*}
\includegraphics[width=175mm]{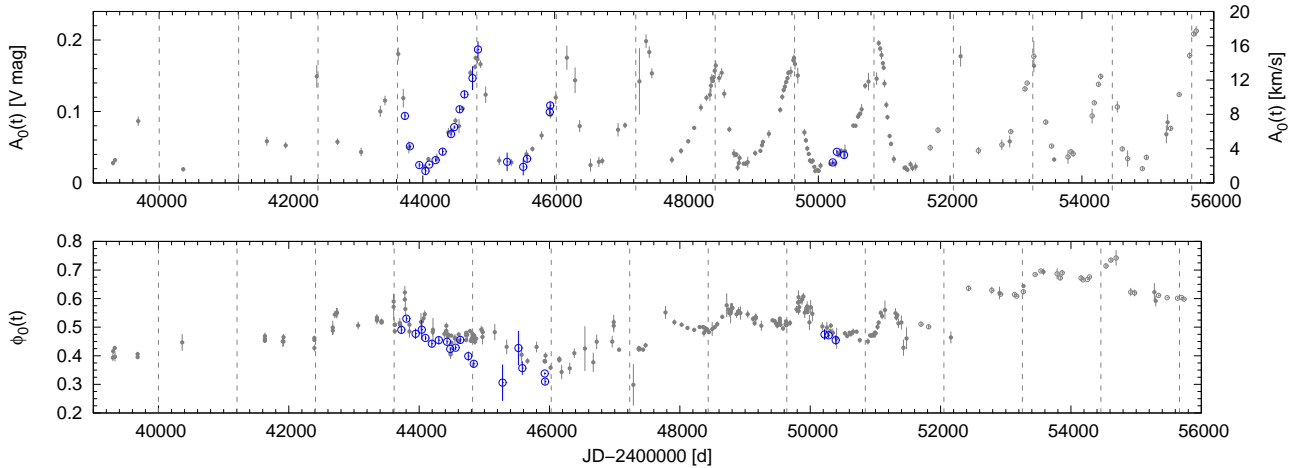}
\caption{Top panel: blue circles represent the amplitude variations 
of the radial velocities, scaled as indicated in 
Sect.~\ref{vradcomp}, grey points are the amplitude variations 
of the photometric data (same as the blue points in 
Fig.~\ref{amp_phase}). Bottom panel: the same for the phase 
variations. Grey vertical lines mark the modulation cycles with 
a spacing of 1205 days. }
\label{vrad_amp} 
\end{figure*}

\subsection{Comparison with the radial velocity}
\label{vradcomp}
We collected a limited amount of RV measurements 
from the literature as well (Fig.~\ref{lc}). We calculated the 
amplitude and phase variations of the data in the same manner 
as we did for the photometry. The results revealed that the 
RV exhibits the same amplitude modulation. 
We calculated the $q=A_{v_{\rm rad}}/A_B$ parameter to confirm 
that the star pulsates in the second overtone (Fig.~\ref{q_parameter}). 
We also scaled the amplitude variations to the \textit{V}-band 
photometric amplitude values in Fig.~\ref{vrad_amp} with a 
scaling factor of $A_{v_{\rm rad}}/A_V=83$. The lower panel of 
Fig.~\ref{vrad_amp} shows that the phase follows the same 
overall slopes as the photometric data but very little data 
are available at the maximum phase peaks. 

In lack of Burki's much more extended RV data, 
we were not able to investigate whether the secondary modulation 
is present in the amplitudes and phases of the RV 
measurements. A visual inspection of Figure~5 in \citet{burki06} 
hints at a slight decrease in the maximum amplitudes from about 
JD~2446500 until JD~2449000, in accordance with our findings. 
However, we cannot exclude that the data set simply misses the 
largest-amplitude pulsation cycles. 

\begin{figure}
\includegraphics[width=84mm]{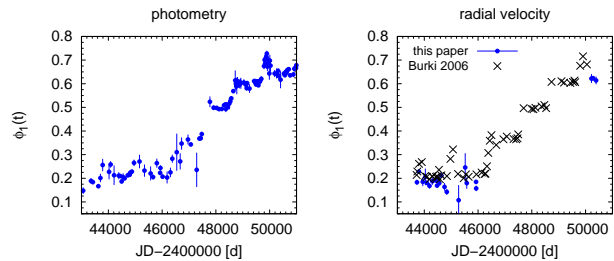}
\caption{Comparison of our phase variation results and the 
RV phase variations obtained by \citet{burki06}. 
Here the basic pulsation frequency is $f=0.670703$ instead of 
the $f=0.670775$ value we used in the rest of the figures. 
Blue points: data used in this paper, grey crosses: data points 
extracted from \citet{burki06}. }
\label{vrad_burki}
\end{figure}

However, we can compare his phase variation results with 
our calculations. The values were extracted from the original 
figure with the DEXTER online 
tool\footnote{http://dc.zah.uni-heidelberg.de/sdexter} so the 
timings are not accurate enough to analyse the data but they make 
a visual comparison possible. \citet{burki06} noted that
the phase seems to experience sudden jumps. If we transform our data 
to the pulsation frequency used in that paper 
($f=0.6707073\, {\rm d}^{-1}$), the similarities become evident. 
As Fig.~\ref{vrad_burki} shows, the apparently flat sections 
of the phase variations are the combined result of the selected 
pulsation period, the shape of the phase modulation curve with the 
sharp peaks at maximum values and the fact that some of these peaks 
are not sampled by \citet{burki06}. This figure provides further 
hints that the secondary modulation is present in the RV variations: 
the change of slope at JD 2446200 coincides with one of the minima 
in the phase variation in Fig.~\ref{phase_longmod}.

\subsection{Comparison with RR~Lyrae stars}
\label{multi}
An important issue is to determine the level of similarity or 
difference to the Blazhko effect observed in RR~Lyrae stars. 
Multiple modulation cycles were identified in several Blazhko stars, 
both from ground- and space-based measurements \citep{czlac,gug12}, 
although the occurrence rate was found to be widely different. 
The \textit{Kepler} sample used by \citet{tailormade} is relatively 
small with only 15 stars, but the ultra-precise photometry revealed 
multiple modulation in 80\% of the sample. Ground-based survey 
data yielded 12\%, based on a larger sample (12/100 stars) but 
with much less precision and temporal coverage, therefore this 
result should be considered as a lower limit \citep{skarka14}. 
In any case, multiple modulation is not unexpected for RR~Lyrae stars.

The primary modulation is relatively slow compared to the 
pulsation period ($P_{m1}/P_{\rm puls}\approx808$). 
However, slow modulation is not uncommon among RR~Lyrae stars either: 
the largest period ratio in the \textit{Kepler} sample is 
$P_{m}/P_{\rm puls}=723/0.5617=1287$ for V454 Lyr (KIC 6183128). 
Stars with modulation periods up to about 3000 days were identified 
in the OGLE-III data \citep{ogle3}.

As Fig.~\ref{folded} shows, the shape of the primary amplitude 
modulation is highly asymmetric. Such modulation envelopes are 
uncommon but not unprecedented among RR~Lyrae stars: \textit{e.g.} 
the stars RX~Col and FS~Vel display a slowly increasing pulsation 
amplitude, followed by a fast decrease towards the minimum amplitude 
\citep{skarka14}.

\begin{figure}
\includegraphics[width=84mm]{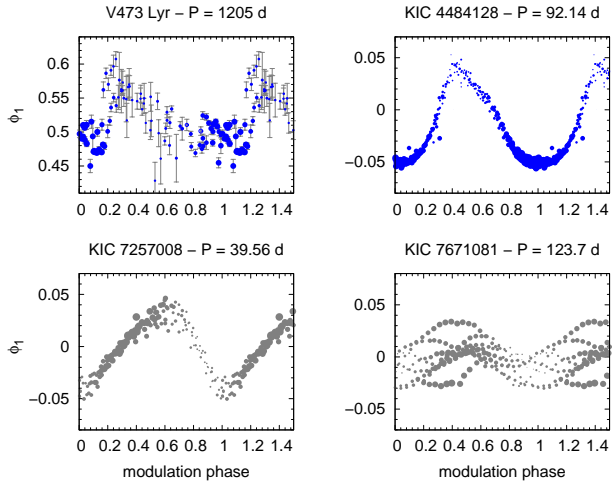}
\caption{Comparison of the amplitude and phase 
modulations between V473~Lyrae and three RR~Lyrae stars. Positions 
of the points indicate the $\phi_1(t)$ value, point sizes the 
corresponding $A_1(t)$ value. All three RR~Lyrae stars are from 
the \textit{Kepler} sample \citep{tailormade}. KIC~4484128 (V808~Cyg) 
shows similar variations while KIC~7257008 and KIC~7671081 (V450~Lyr) 
display very different curves. Modulation periods are indicated.}
\label{amp_phase_rel}
\end{figure}

Another possibility for comparison is the phase relation between 
the pulsation amplitude and phase variations. Blazhko RR~Lyrae 
stars show a wide variety of modulation relations. If plotted as a 
loop diagram (e.g. $\phi_1(t)$ plotted against $A_1(t)$), the 
shape of the loops and the direction of progression indicate the 
relation of the two measurements. We presented a loop diagram for 
V473~Lyrae in our previous paper \citep{v473first}. 
The visualization of the loops of the star, however, required 
additional smoothing. Here we follow a different method and plot 
the phase variations instead, folded by the (primary) modulation 
period. The corresponding pulsation amplitude values are then 
indicated with the sizes of the points (Fig.~\ref{amp_phase_rel}). 

We selected three RR~Lyrae stars from the \textit{Kepler} sample 
for comparison \citep{tailormade}. Two stars (KIC~7257008 and 
KIC~7671081, bottom row) display very different modulations. 
In the case of the former star, a roughly quarter-period shift is 
evident between the amplitude and phase, \textit{i.e.}\ maximum 
and minimum amplitudes occur around the middle of the rising and 
descending branches of the phase variations. In the latter case, 
the relation between the amplitude and phase modulations themselves 
varies over time. The third star (KIC~4484128, V808~Cyg), however, 
shows variations similar to V473~Lyr (top row) with antiparallel 
phase and amplitude variations (which translates to parallel amplitude 
and $O-C$ variations). These examples, together with the asymmetry 
and length of the primary modulation and the presence of the 
secondary modulation indicate that analogues to V473~Lyrae can be 
found among the RR~Lyrae stars. We must point out, however, that 
they also indicate that the variations collectively referred to as 
the Blazkho effect itself is a very diverse phenomenon.

\section{Hydrodynamic models}
Currently the most plausible model of the Blazhko effect is the 
mode resonance hypothesis \citep{bk11}. Given that the pulsation 
of RR~Lyrae stars and Cepheids share many similarities, one can 
expect that mode resonances occur in Cepheids too. To find out, 
we calculated a large set of linear hydrodynamic models with the 
Florida-Budapest code \citep{kollath01,kollath02}. 
Initial fundamental parameters (effective temperature, mass, 
luminosity) were based on earlier results from the literature.

\citet*{fabregat} derived these parameters from $uvby\beta$ photometric 
observations and found that 
$T_{\rm eff}=5860 \pm 100\,K$, $L=960\pm390\,L_\odot$ and 
$M=4.1 \pm 2.5 \,M_\odot$. Based on spectroscopic observations and 
comparison with previous studies, \citet{andrievsky} determined the 
parameters to be $T_{\rm eff}=6100\,K$, $L=740\,L_\odot$ with 
pulsational and evolutionary masses of $M_{\rm puls}= 3.0\, M_\odot$ 
and $M_{\rm evol}= 4.6\, M_\odot$. Therefore our model grid was set up 
with the following limits and steps: $M=4 - 5\, M_\odot$, $0.1\, M_\odot$; 
$T_{\rm eff}=5800 - 6400\, K$, $25 \,K$; $L=720 - 950\,L_\odot$, 
$5\, L_\odot$ and a chemical composition of $X=0.70$; and $Z=0.016$. 

\begin{figure}
\includegraphics[width=84mm]{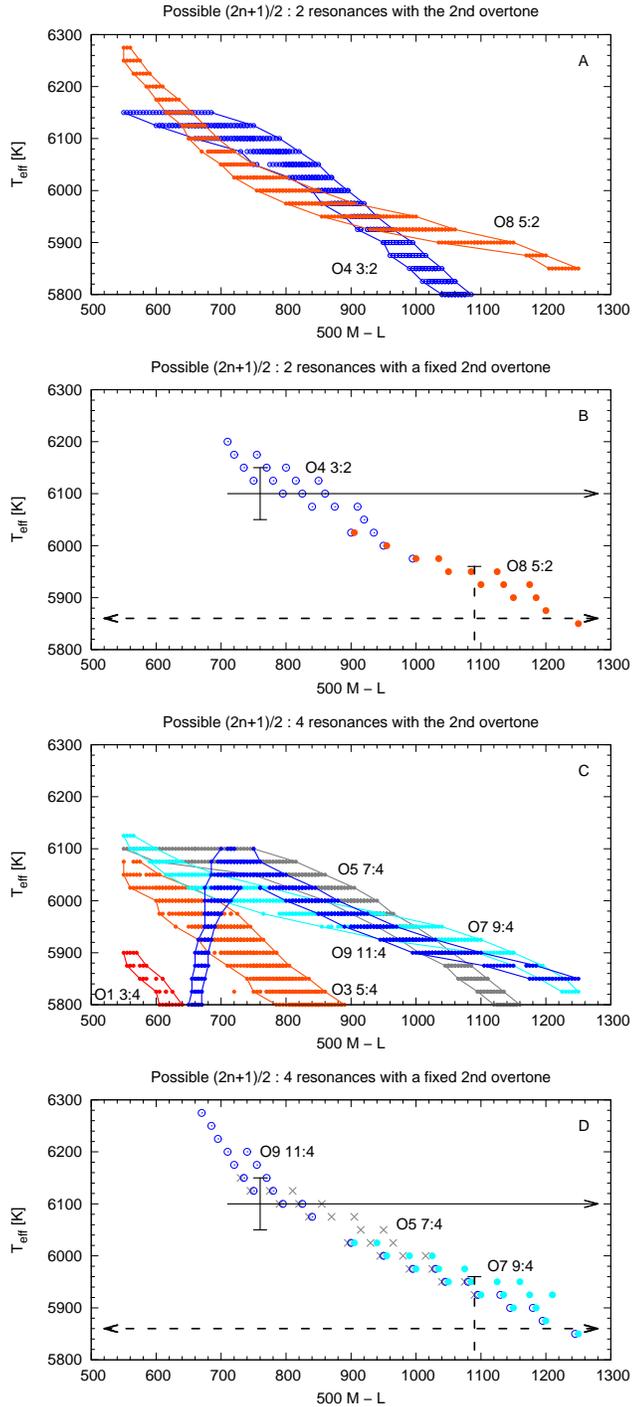}
\caption{The diagnostic diagram of the linear hydrodynamic Cepheid 
models. Resonance regions are plotted in a 2D projections of the 
original $T_{\rm eff} - M - L$ space. Mass and luminosity values are 
in solar units. The black crosses and arrows indicate the possible 
positions of V473~Lyrae. A: half-integer resonances with the second 
overtone. B: solutions close to half-integer resonances with the 
second overtone fixed at $P=1.4909$ d. C: quarter-integer resonances. 
D: solutions close to quarter-integer resonances with a fixed 
overtone period. }
\label{diagn}
\end{figure}

We searched the model grid for resonances between the linear periods 
of the second overtone and other radial modes. The results are
summarized in Fig.~\ref{diagn}. The plots are similar diagnostic 
diagrams that were created for RR~Lyrae models by \citet{kmsz11}. 
The resonant solutions spread out to a relatively flat surface in 
the three-dimensional $T_{\rm eff}-M-L$ parameter space. A well-chosen 
2D projection of this space reduces the subspace of resonant models 
into a relatively thin strip thus making the visualization much easier. 
For RR~Lyrae stars, the projection was 
$T_{\rm eff}$ \textit{vs.}\ $150\,M-L$ where \textit{M} and \textit{L} 
are in solar units. For the much more luminous Cepheids, the 
projection was found to be $500\,M-L$, in solar units (Fig.~\ref{diagn}). 

\subsection{Linear resonances}
A main difference from the RR~Lyrae study of \citet*{kmsz11} is that 
V473~Lyr is pulsating in the second overtone instead of the fundamental 
mode. Therefore period ratios with the higher modes are very 
different. \citet{ak97} noted that the $P_2/P_6=2$ resonance can occur with the sixth overtone in their radiative nonlinear second-overtone models. However, \citet{mb90} showed that in W~Vir models, only the half-integer resonances destabilize the pulsation mode, and the same was observed in RR~Lyrae models \citet*{kmsz11}. For the RR~Lyrae models several half-integer $((2n+1)/2)$ 
resonances that occur with a whole spectrum of overtones were identified. 
In contrast, we could find only two cases for the second-overtone 
Cepheid models, as seen in panel A of Fig.~\ref{diagn}: a 3:2 
resonance with the 4th overtone and a 5:2 resonance with the 8th 
overtone. In the plot we selected models with period ratios that 
are within $\pm 0.1\%$ from the exact resonance value. 

From the two cases, the 8th overtone is more interesting: 
it may be a strange mode, \textit{i.e.}\ it can be trapped in the 
uppermost layers of the star and alter the pulsation significantly. 
Such interaction was recently observed in RR~Lyrae stars where a 
resonance between a strange mode and the fundamental mode causes a 
period doubling bifurcation in the pulsation \citep*{pd,kmsz11}.

Although the 8th overtone is a good candidate for the mode-resonance 
induced instability of the pulsation, we looked for other resonances 
too. Panel C in Fig.~\ref{diagn} shows the locations of the 
quarter-integer $((2n+1)/4)$ resonances. In this case, we identified 
five possibilities: resonances with 1st, 3rd, 5th, 7th and 9th overtones. 

Focusing now on V473~Lyrae, there is an additional constraint to the 
models: the pulsation period. In panels B and D we only plotted the 
solutions that are both close to the resonances and close to the 
period of the second overtone ($P_2=1.485-1.492$ d). Since nonlinear 
models can lock into resonances even if the period ratios of the 
linear models are slighly off, we extended the allowed period ratio 
range to  $\pm 0.5\%$. The fundamental parameters of \citet*{fabregat} 
and \citet{andrievsky} are also indicated with dashed and continuous 
lines, respectively. In the former case for the mass we used the 
$M_{\rm puls}= 3.0\, M_\odot$ value. The two positions of the star 
align well with the selected models but select different resonances. 

However, there are complications to these results. Large uncertainties 
in the luminosity and especially in the mass of the star make the 
$500M-L$ parameter very vague. The experiences with RR~Lyrae models 
also tell us that metallicity can shift the position of the resonant 
solutions in the diagram considerably \citep*{kmsz11}. 
The difference between the linear and nonlinear mode frequencies 
also shifts the region of nonlinear resonant models compared to the 
linear ones. Therefore the linear models only tell us that there 
indeed are plausible resonances to drive the modulation but only 
further research will be able to reveal which, if any, may be operating 
in V473~Lyrae. The effects of different metallicities and the analysis of the nonlinear models will be detailed in a separate paper.

\section{Conclusions and future work}
We examined the photometric data of the Cepheid 
V473~Lyrae collected by various observers and covering about 44 years 
in several photometric bands. The star shows strong amplitude and phase 
modulations with an average period of $P_{m1} = 1205 \pm 3$ days. 
With the help of the limited amount of RV measurements 
at hand, we also confirmed that the star is pulsating in the second 
overtone. We identified modulation side-lobes up to the third order 
(septuplets).

It turned out, however, that the light variations of the star are 
even more complex. After a careful analysis we identified a secondary 
modulation both in the Fourier spectra and the amplitude and phase
variations of the star. This tertiary period of the star is in 
the range of $P_{m2}= 5300 \pm 150$ days, or 14.5 years and detectable 
in both the amplitude and phase variations. Recent results revealed 
that multiple modulations not only occur \citep{skarka14} but they 
may prevail in Blazhko stars \citep{tailormade}. We also detected that 
the length of the primary modulation cycle changes in parallel with 
the secondary cycle. We expect that the two cycles interact with 
each other but the current amount of data is insufficient to decide 
this question.

Phenomenologically, we consider the light curve variations of 
V473~Lyrae equivalent with the Blazhko effect. Although some details, 
such as  the asymmetry of the primary modulation or the relation 
of the amplitude and phase variations seem to differ from the general 
picture, we can find an RR~Lyrae counterpart for each particular 
aspect. This fact underlines that the Blazhko effect is a very 
diverse phenomenon and without physical models, the similarities 
remain convincing but not conclusive.

There are several open questions that deserve further attention. 
These ground-based data lack the precision to identify low-amplitude 
additional modes that may drive mode interactions in the star. 
In RR~Lyrae stars, space-based photometry revealed that almost 
all modulated stars show additional frequency components in the 
Fourier spectra at the millimag level \citep{benko10}. Continuous 
coverage was crucial to the discovery of period doubling \citep{pd}. 
Currently the only space-based photometer that can observe V473~Lyr 
is the Canadian MOST satellite \citep{most}. We successfully applied 
for telescope time and MOST will observe the star for one month 
in July 2014. Although one month is a very short time compared 
to the modulation cycles, the continuous and precise monitoring 
of the star will be essential to identify additional modes and 
cycle-to-cycle variations (assuming that they will be present 
during those $\sim 20$ pulsation cycles). The TESS space 
telescope\footnote{http://space.mit.edu/TESS}, to be launched 
in 2017, will also observe the star for one month.

Continued follow-up of the modulation cycles is also important. 
On the one hand, our analysis was limited by the strongly varying 
temporal coverage and photometric precision, the use of different 
filters and the systematic differences that are inherent to the 
data combined from several sources. On the other hand, the star 
is too bright for most automated survey telescopes. A new initiative, 
however, may provide extended, homogeneous measurements of the star. 
The Fly's Eye\footnote{http://www.flyseye.net} project will use 19 
small, wide-field cameras to provide a high-cadence, multicolour 
time-domain survey of the entire visible sky down to 30 degrees 
altitude \citep{fly1}. The prototype of the system may start 
observing the star from this year on. We also encourage amateur 
astronomers to follow the star's light variations with digital photometry.

Finally, non-linear hydrodynamic modelling of second-overtone 
Cepheids will be vitally important to determine whether mode 
resonances occur in these stars. Recent advancements in 1D pulsation 
codes revealed that both RR~Lyrae and BL~Herculis (short period Type~II  
Cepheid) models may exhibit multimode pulsation, mode resonances, chaos, 
and small-amplitude modulation \citep*{kmsz11, models, plachy13, sm12, sm14}. 
If mode interactions can be identified in second-overtone Cepheid models, 
they will undoubtedly support the case of the mode resonance model, 
and will point towards a unified model for the Blazhko effect in Cepheid 
and RR~Lyrae stars.

\section*{Acknowledgements}
We thank the referee, Shashi Kanbur, for his comments that helped to 
improve the paper. Fruitful discussions with J\'ozsef Benk\H{o} are 
gratefully acknowledged. The work of L. Moln\'ar leading to this research 
was supported by the European Union and the State of Hungary, 
co-financed by the European Social Fund in the framework of 
T\'AMOP 4.2.4. A/2-11-1-2012-0001 `National Excellence Program'. 
This work has been supported by the Hungarian OTKA grant K83790, 
the ESTEC Contract No.\ 4000106398/12/NL/KML, and the `Lend\"ulet-2009' 
Young Researchers' Programme of the Hungarian Academy of Sciences. 
We acknowledge the AAVSO International Database contributed by 
observers worldwide.

\label{lastpage}

\end{document}